# Recent advances in the determination of some Galactic constants in the Milky Way

Jacques P. Vallée

National Research Council of Canada, National Science Infrastructure, Herzberg Astronomy & Astrophysics, 5071 West Saanich Road, Victoria, B.C., Canada V9E 2E7

Keywords: Galactic structure; Galactic constants; Galactic disk; statistics

**Abstract.**
Here we statistically evaluate recent advances in determining the Sun-Galactic Center distance ($R_{sun}$) as well as recent measures of the orbital velocity around the Galactic Center ($V_{lsr}$), and the angular rotation parameters of various objects. Recent statistical results point to $R_{sun} = 8.0 \pm 0.2$ kpc, $V_{lsr} = 230 \pm 3$ km/s, and angular rotation at the Sun (ω) near $29 \pm 1$ km/s/kpc for the gas and stars at the Local Standard of Rest, and near $23 \pm 2$ km/s/kpc for the spiral pattern itself.

This angular difference is similar to what had been predicted by density wave models, along with the observation that the galactic longitude of each spiral arm tracer (dust, cold CO) for each spiral arm becomes reversed across the Galactic Meridian (Vallée 2016b).

## 1. Introduction

The value of the distance of the Sun to the Galactic Center, $R_{sun}$, is one of the fundamental parameters of our Milky Way disk galaxy. Similarly, the value of the circular orbital velocity $V_{lsr}$, for the Local Standard of Rest at radius $R_{sun}$ around the Galactic Center, is another fundamental parameter. In 1985, the IAU recommended the use of $R_{sun} = 8.5$ kpc and $V_{lsr} = 220$ km/s. Recent measurements of $R_{sun}$ and $V_{lsr}$ deviate somewhat from this IAU recommendation.

Section 2 makes a table of recent measurements of two galactic constants, namely the distance of the Sun to the Galactic Center, and the orbital circular velocity near the Sun around the Galactic Center. Section 3 deals with current angular rotation rates for various galactic features, while Section 4 concludes.

## 2. Galactic constants $R_{sun}$ and $V_{lsr}$

**Table 1** shows about forty *very recent* measurements of $R_{sun}$ and/or $V_{lsr}$, as published from mid-2012 to early-2017. Each different type of measurement (column 3) comes with its own assumptions, and each assumption comes with its own error.

As shown statistically in the last rows of Table 1, one finds here that the median and unweighted mean for $R_{sun}$ are close to $8.0 \pm 0.2$ kpc, while the median and mean for $V_{lsr}$ are close to 230 ± 3 km/s. Next, we show a weighting using the original authors' own error bars, without personally interjecting an opinion, Thus we then chose a weight (inversely proportional to the stated error) for each published data, giving a mean for $R_{sun}$ of 8.0 ±0.2 kpc, and for $V_{lsr}$ of 228 ± 2 km/s.

All recent reviews since 2010 gave values for $R_{sun}$ below the IAU-recommended value of 8.5 kpc. Each review covers a different time period (none included published data since early 2016) and each weights different methods differently (some earlier published results here were not included in some reviews). Malkin (2013) found a weighted mean value of 8.0 ± 0.4 kpc. Gillessen et al (2013- their Fig.2) showed a median near 8.1 ±0.3 kpc. Bland-Hawthorn & Gerhard (2016) had an unweighted arithmetic mean of 8.0 ±0.4 kpc (their Table 3) and a weighted mean of 8.2 ±0.1 kpc (their Fig. 4a). De Grijs & Bono (2016) yielded $R_{sun}$ = 8.3 ±0.4 kpc.

Most measurements in Table 1 are for either $R_{sun}$ or $V_{lsr}$, but rarely together.
**Figure 1** shows a plot of these 9 cases of dual measurements, and the weighted fit of a line shown dashed. These dual measurements gave an increasing $V_{lsr}$ with an increase in $R_{sun}$, with the Sun's distance at 8.0 kpc and the LSR's velocity at 230 km/s.

Figure 2 shows a histogram of the 27 most recent observational determinations of $R_{sun}$ from Table 1, showing a spread of about 3 kpc and a peak near 8 kpc.

Figure 3 shows a histogram of the 24 most recent observational determinations of $V_{lsr}$ from Table 1, showing a spread of about 40 km/s and a peak near 230 km/s.

3. **Galactic angular rotation parameter**

As gas and stars follow their orbit around the Galactic Center, the quantitative value of the angular rotation of the Local Standard of Rest [LSR] can be obtained from the mean values in Table 1, along with the orbital period around the Galactic Center.

Using 8.0 and 230 km/s for the LSR's constants, the ratio $V_{lsr} / R_{lsr} = \omega$, and $\omega$ is the angular orbital rotation around the Galactic Center. Here $\omega$ = 28.8 km/s/kpc. This angular rotation value is close to that found earlier by Nagayama et al (2011).

In a flat rotation curve, the Oort's constant A = +0.5 V/R = +14.40 km/s/kpc, and the Oort's constant B = -0.5 V/R = -14.40 km/s/kpc, while the angular rotation = A – B = 28.8 km/s/kpc.

**Table 2** assembles some data on these angular rotations and rotation periods, as published recently.

Figure 4 shows a diagram of most features mentioned in Table 2. The 'spiral pattern' sits on top of the observed locations of arms (stars and gas); this pattern goes around the Milky Way at a slightly slower speed than that of the orbiting gas and stars (which cross the spiral pattern). There is a smaller angular velocity for the spiral arm pattern (23 km/s/kpc) than for the gas and stars (29 km/s/kpc).

The dust lane is located on the inner side of each spiral arm, closest to the Galactic Center. The galactic longitudes of each spiral arm tracer (dust, cold CO) for each spiral arm shows a reversal on both sides of the galactic Meridian (Vallée 2016b), similar to what had been predicted by density wave models.

The angular rotation for the spiral pattern is near ($\approx 23 \pm 2$ km/s/kpc) in the Milky Way Galaxy, which is similar to that found elsewhere in the fairly isolated galaxies NGC 6946 ($\approx 22$ km/s/kpc) and NGC 2997 ($\approx 16$ km/s/kpc); it differs from that for the strongly interacting galaxy M51 ($\approx 38$ km/s/kpc) – for a review see Ghosh & Jog (2016 – table 2).

It is more difficult to obtain observational values for the angular rotation and the rotation period of the boxy bulge bar extending to 2.1 kpc, as well as for the thin long bar extending to 4.2 kpc (see Table 3 in Vallée 2016a).

The angular rotations for the spiral arm pattern, and for the two bars, are model-deduced from other observed parameters, thus not directly observed, and not known to better than a few km/s/kpc.

Some models (e.g., Bissantz et al 2003) required two different angular speeds, one for the bar and another one for the spiral pattern, sufficiently *apart* to prevent a dissolution of the spiral arms. Other models (e.g., Englmaier & Gerhard 1999) have argued that the angular rotation of the spiral pattern ought to be similar to the angular rotation of the short boxy bulge bar, if the two features are *attached*. Given that the difference in Table 2 between these two angular rotations ($\approx 12$ km/s/kpc) has a quadratically combined error of 12 km/s/kpc, one cannot chose a model in particular (apart or attached).

Some other models (e.g., Li et al 2016) have argued that the angular rotation of the short boxy bulge bar ($\approx 35 \pm 10$ km/s/kpc) ought to be similar to the angular rotation of the long thin bar ($\approx 46 \pm 15$ km/s/kpc), if the two bars are dynamically *stable*. Recently, Monari et al (2017) suggested it to be a *loosely* wound spiral structure, with its own rotation. Given that the difference between these two angular rotations ($\approx 11$ km/s/kpc) has a quadratically combined error of 18 km/s/kpc, one cannot chose a model in particular (stable or loose).

## 4. Conclusion

We assembled recent measurements of two basic galactic constants (Table 1). We find $R_{sun}$ close to 8.0 kpc, while $V_{lsr}$ is close to 230 km/s. When both the LSR circular orbital velocity and the distance of the Sun to the Galactic Center are measured in the same experiments, over a wide range of methods, the results display a trend of increasing $V_{lsr}$ with an increase of $R_{sun}$ values (Figure 1).

Recent studies of the angular rotation rates of some features in the Milky Way (Table 2) are assembled, combining data from Table 1 here and from the literature.

At this time, it is difficult to choose among bar models. More accurate observational data are needed to separate among bar models and spiral pattern.

**Acknowledgements.**
The figure production made use of the PGPLOT software at NRC Canada in Victoria. I thank an anonymous referee for useful, careful, and historical suggestions.

**Table 1 – Recent measures of global parameters of the Milky Way since mid-2012**

| $R_{sun}$ (kpc) | $V_{lsr}$ [a] (km/s) | Data used | References |
|---|---|---|---|
| 8.0±0.8 | 218±6 | 3365 stars | Bovy et al (2012 – tab. 2) |
| 7.7±0.4 | - | S0 around Gal.Cntr. | Morris et al (2012 – sect.5) |
| 8.0±0.4 | - | VLBI astrometry | Honma et al (2012 – tab.7) |
| 8.3±0.4 | 238±9 | SLOAN stars | Schonrich (2012 – sect 4.3) |
| 7.6±0.3 | 217±11 | Cepheids near Sun | Bobylev (2013 – tab.3) |
| 8.2±0.8 | - | O-B5 stars | Zhu & Shen (2013 – tab. 1) |
| 8.0±0.7 | - | Open clusters | Zhu & Shen (2013 – tab. 1) |
| 8.0±0.8 | - | Classical cepheids | Zhu & Shen (2013 – tab. 1) |
| - | 239±16 | 73 masers | Bobylev & Bajkova (2013 – sect.4) |
| - | 234±5 | 58 masers | Bajkova & Bobylev (2013 – sect.5) |
| - | 205±15 | 4400 C stars | Battinelli et al (2013 – fig.3) |
| 8.5±0.4 | - | nuclear star cluster,S2 | Do et al (2013 – sect. 6) |
| 8.3±0.2 | - | RR Lyrae stars | Dékany et al (2013 – sect. 4) |
| 7.6±0.6 | - | Type II Cepheids | Matsunaga et al (2013 – fig.10) |
| 8.2±0.2 | - | Red Giant clumps | Cao et al (2013 – tab. 4) |
| 8.3±0.2 | 240±8 | 80 masers | Reid et al (2014 – sect. 4.4) |
| 6.7±0.4 | 203±12 | OB stars | Branham (2014 – tab.3) |
| 7.5±0.3 | - | red clump stars | Francis & Anderson (2014 –sect.7) |
| 7.4±0.3 | - | 154 globular clusters | Francis & Anderson (2014 –sect.3) |
| - | 233±2 | RAVE stars | Sharma et al (2014 – table 11) |
| 8.3±0.3 | 233±13 | Palomar 5 glob. clusters | Kupper et al (2015 – Sect. 4.1.1) |
| 7.7±0.1 | - | 36 061 A - F stars | Branham (2015 – Sect.5) |
| 8.3±0.1 | - | nuclear star cluster | Chatzopoulos et al (2015 – sect. 4.2) |
| 8.3±0.4 | - | RR Lyrae stars | Pietrukowicz et al (2015 - sect. 4.2) |
| 8.0±0.3 | - | 93 masers | Bajkova & Bobylev (2015 – sect.4) |
| - | 230±10 | 119 masers | Bobylev & Bajkova (2015a – fig.1) |
| - | 234±14 | 120 spectr. binaries | Bobylev & Bajkova (2015b- fig.2) |
| - | 225±10 | open clusters,OB assoc. | Melnik et al (2016 – fig. 10) |
| - | 228±14 | GMC, CO J=1-0 gas | McGaugh (2016 – table 1) |
| 8.9±0.4 | - | 4883 MIRA stars | Catchpole et al (2016 – sect. 7) |
| - | 210±10 | thin disk stars | Rojas-Arriagada et al (2016 – Fig.8) |
| - | 240±6 | 16000 red clump stars | Huang et al (2016 – Fig.6) |
| - | 230±12 | 183000 RAVE4 stars | Bobylev and Bajlkova (2016 – Sect.5) |
| 7.9±0.1 | - | S0 orbit astrometry | Boehle et al (2016 – Table 4) |
| - | 240±8 | HI emission | Reid & Dame (2016 – Fig.4) |
| - | 236±6 | Open star clusters | Bobylev et al (2016 – fig.2) |
| - | 231±6 | Cepheids | Bobylev (2017 – Fig.2) |
| - | 219±8 | OB stars | Bobylev & Bajkova (2017 – Sect.8) |
| 8.2±0.1 | 233±3 | mass model (main) | McMillan (2017 – Table 2) |
| 8.0±0.2 | 227±4 | mass model (alternate) | McMillan (2017 – Table 6) |
| 7.6±0.1 | - | GKM stars | Branham (2017 – Table 3) |
| 8.4±0.1 | 243±10 | 131 masers | Rastorguev et al (2017 – tab.1) |

- - - - - - - - - - - - - - - - - - - - - - - - - - - - - - - - - - - -

| | | |
|---|---|---|
| **8.0** | **232** | **Median value (all data)** |
| **8.0 ±0.2** | **229±3** | **Mean value (unweighted; all data), and standard dev. of the mean** |
| **8.0 ±0.2** | **228±2** | **Mean value (weighted; all data), and standard dev. of the mean** |
| **8.0** | **230** | **Adopted here (within one s.d.m. of mean value)** |

- - - - - - - - - - - - - - - - - - - - - - - - - - - - - - - - - - - -

Note:
 (a): Local Standard of Rest - the kinematic circular rotation value near the Sun, excluding the Sun's peculiar velocity.

**Table 2 – Recent studies of the angular rotation of some features of the Milky Way**

| Feature | R Gal. radius (kpc) | ref.(a) | ω ang. vel. (km/s/kpc) | ref.(a,b) | V circ. vel. (km/s) | ref(a, b) | P orbital period (Myr) | ref.(a,b) |
|---|---|---|---|---|---|---|---|---|
| LSR (c) | 8.0 | Tab.1 set | 28.8 | eq.1 | 230 | Tab.1 | 220 | eq.2 |
|  | .. |  | 28.7 | N11 | 230 | eq.1 | 219 | eq.2 |
| Spiral pattern | 8.0 | set | 25 | D15; G11 | 200 | eq.1 | 250 | eq.2 |
|  | .. | .. | 23 | L16, J15 | 184 | eq.1 | 273 | eq,2 |
|  | .. | .. | 20 | K16 | 160 | eq.1 | 314 | eq.2 |
| Hot Halo gas | 8.0 | set | 23 | eq.1 | 183 | H16 | 275 | eq.2 |
| Boxy bulge bar | 2.1 | V16a | 30 | R08 | 63 | eq.1 | 209 | eq.2 |
|  | .. | .. | 40 | Q15 | 84 | eq.1 | 157 | eq.2 |
| Thin long bar | 4.2 | V16a | 59 | G11 | 248 | eq.1 | 106 | eq.2 |
|  | .. | .. | 33 | L16a | 139 | eq.1 | 190 | eq.2 |

Notes:
(a): Literature cited: D15 = Dambis et al (2015); G11 = Gerhard (2011 – Sect.4); H16 = Hodges-Kluck et al 2016; J15 = Junqueira et al (2015 – tab.4); K16 = Koda et al (2016 – Tab.2); L16 = Li et al (2016 – sect. 2.2.3); L16a = Li et al (2016 – sect. 4.5); N11 = Nagayama et al (2011 – fig. 6b); Q15 = Qin et al (2015 – Sect. 2); R08 = Rodriguez-Fernandez & Combes (2008 – Sect. 6.1); V16a = Vallée (2016a – tab.3)

(b): Equation 1 :  $R_{(kpc)} \cdot \omega_{(km/s/kpc)} \approx V_{(km/s)}$
   Equation 2 :  $2 \cdot \pi \cdot R_{(pc)} \approx V_{(km/s)} \cdot P_{(Myr)}$
   with:   1 Myr = 3.16 x $10^{13}$ s
   and:    1 pc = 3.09 x $10^{13}$ km

(c) : LSR = Local Standard of Rest, surrounding the Sun's position in the Milky Way disk.

**Figure captions**

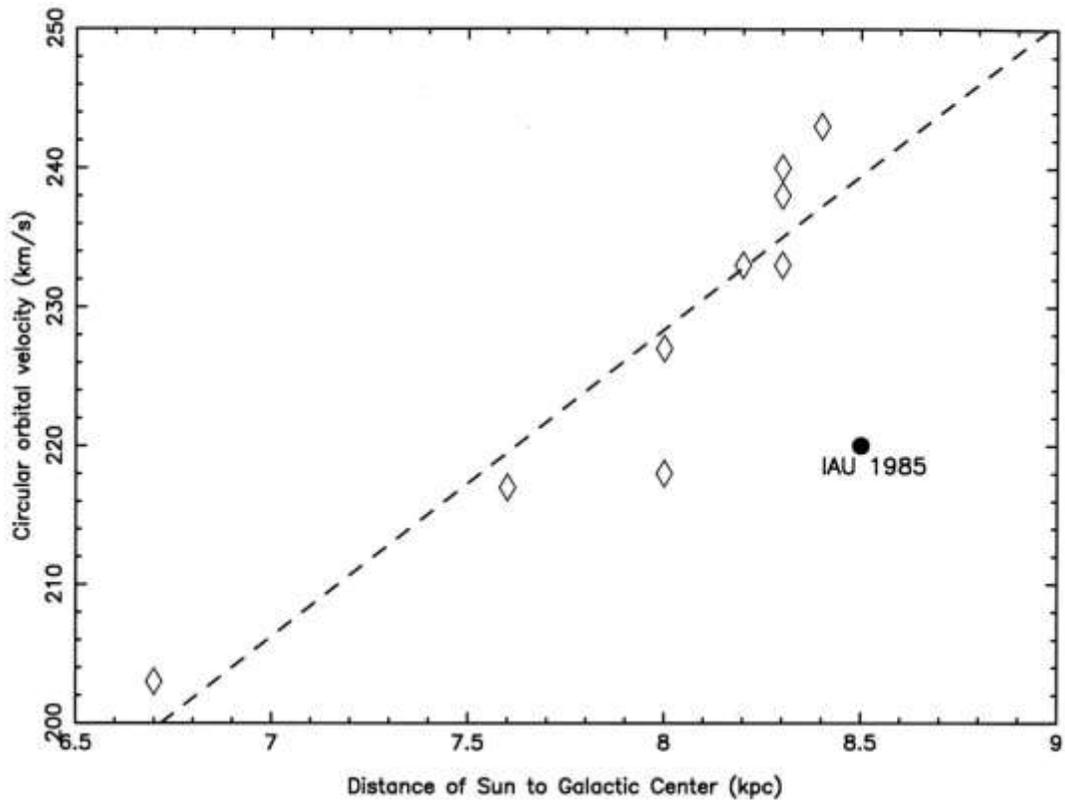

Figure 1. Plot of nine observations that measured both the solar distance to the Galactic center (horizontal) and the LSR circular orbital rotation velocity (vertical), as taken from Table 1. The dashed line is a weighted fit to the nine data in this figure.

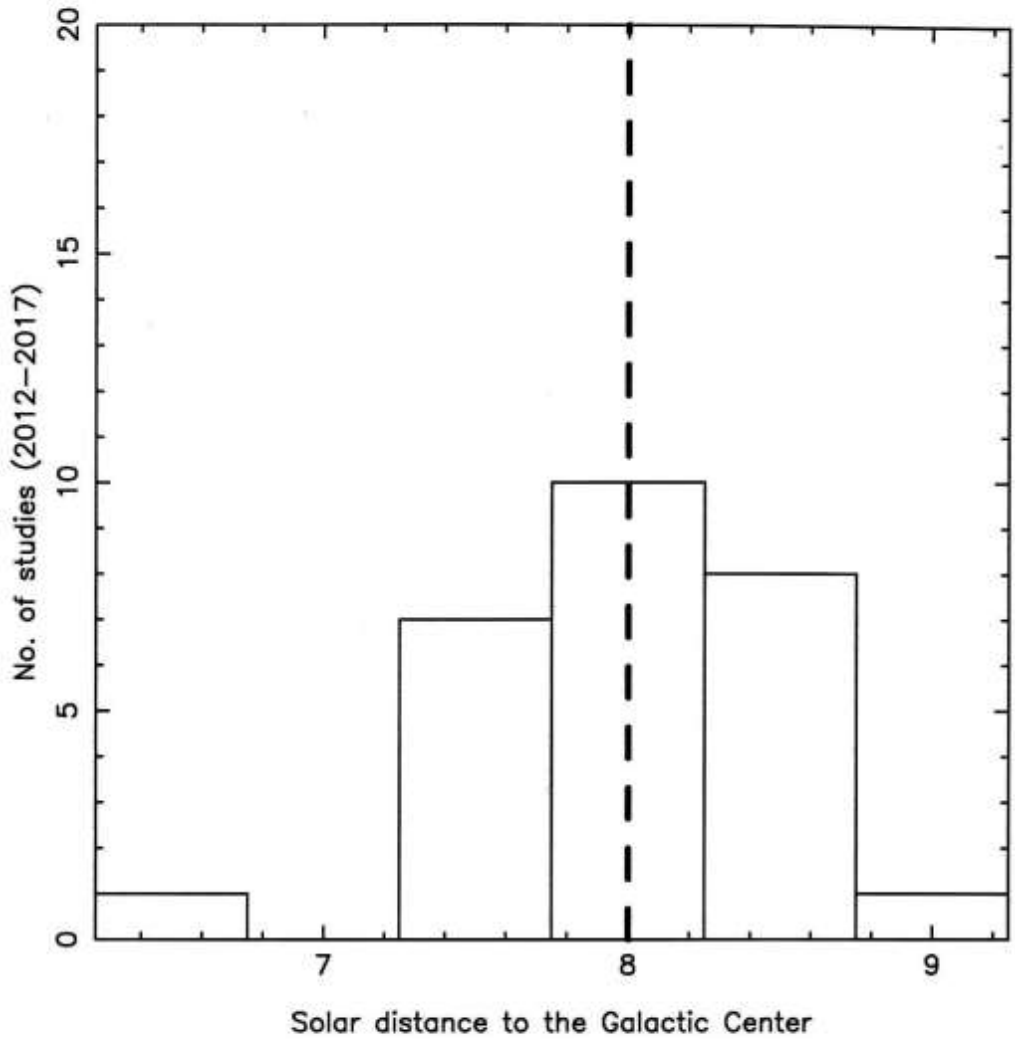

Figure 2. Histogram of the 27 data in Table 1 on $R_{sun}$. Here the spread in recent measurements is from 6 to 9 kpc (x-axis), with a peak near 8 kpc.

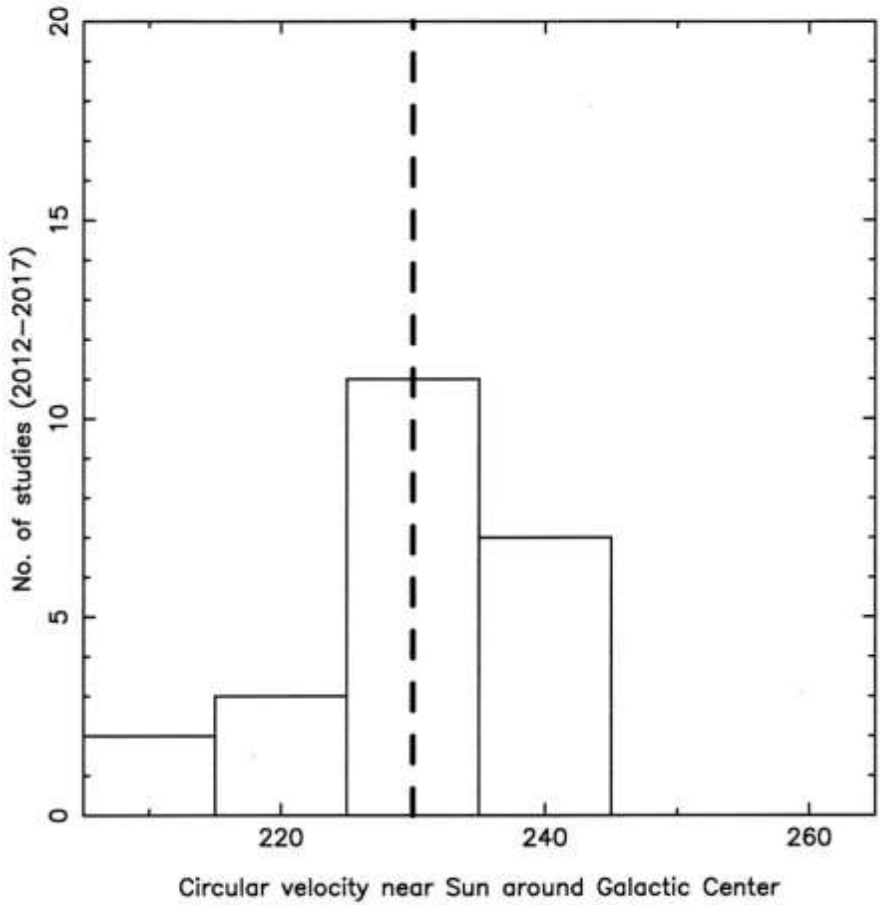

Figure 3. Histogram of the 24 data in Table 1 on $V_{lsr}$. Here the spread in recent measurements is from 205 to 245 km/s (x-axis), with a peak near 230 km/s.

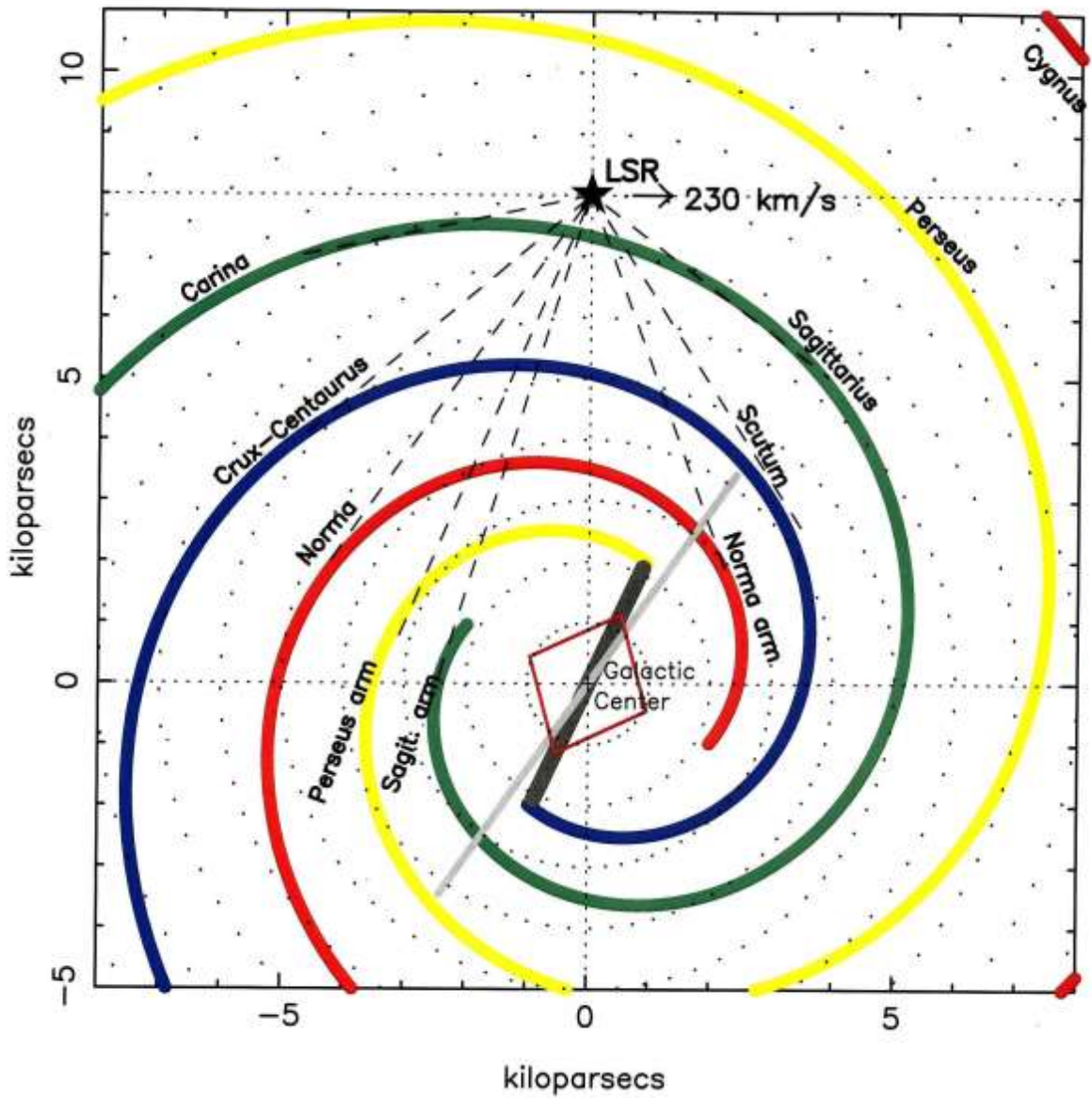

Figure 4. Diagram of the Milky Way's disk plane, showing some of the various features in Table 2. The Local Standard of Rest [LSR], shown by a star, and its direction (arrow) and velocity (230 km/s) are indicated. The LSR englobes many local stars, including the Sun, and has an angular velocity near 29 km/s/kpc. The observed spiral arms are shown, located on the theoretical 'spiral pattern', with an angular velocity near 23 km/s/kpc. Near the Galactic Center, the boxy bulge bar (dark gray) and the thin long bar (light gray) are shown. The hot halo gas (not shown) stand above and below the disk plane. Observed tangents to the spiral arms are shown as dashes. Circles at different radii from the Galactic Center are shown as dots.